\documentclass[epjST]{svjour}
\usepackage{graphicx}
\usepackage{bm}
\begin{document}
\title{Infinite-range transverse field Ising models and quantum computation}
\author{Jun-ichi Inoue\thanks{\email{j\_inoue@complex.ist.hokudai.ac.jp}}}
\institute{Graduate School of Information Science and Technology, Hokkaido University, N14-W9, Kita-ku, 
Sapporo 064-0814, Japan}
\abstract{
We present a brief review on 
information processing, computing and inference via quantum fluctuation, 
and clarify the relationship between the probabilistic 
information processing and theory of quantum spin glasses 
through the analysis of the infinite-range model.  
We also argue several issues to be solved for the future direction in the research field. 
} 
\maketitle
\section{Introduction}
\label{intro}
Recently, 
a new paradigm of 
quantum computation based on quantum-mechanical fluctuation
has drawn a lot of attentions of researchers who are working in 
not only physics but also interdisciplinary fields such as computer sciences 
or system engineering.    
Especially, quantum fluctuations by means of the transverse field \cite{Suzuki} have been 
investigated extensively within the context of combinatorial optimization problems, 
which induce quantum-mechanical tunneling instead of thermal jumps between possible candidates of the solution 
\cite{Kadowaki,Farhi,Santoro}. 
Thus, the algorithm is called as {\it quantum annealing }(QA) or  {\it quantum adiabatic algorithm}. 
The QA has been 
applied to various optimization problems by 
solving the Schr\"{o}dinger equation 
or carrying out quantum Monte Carlo simulations on classical computers. 

Besides purely theoretic studies or 
conceptual discussions,  
quantum algorithm by means of the QA 
has been implemented in hard ware by D-wave systems based in British Columbia 
\cite{Harris,Johnson,Boixio}, 
and several positive reports on the project have been released.  
Taking into account these scientific and technological developments,  
quantum fluctuations induced by transverse fields could have the potential to provide us several effective tools for solving combinatorial optimization problems. 

Apparently, 
one of the motivations of QA  
is similarity 
between 
thermal and quantum fluctuations. 
The former is now recognized as a well-established 
optimization 
method called as {\it simulated annealing} (SA).  
By controlling the temperature $T$, which 
is defined as a degree of standard deviation of 
physical quantity $X$ as 
\begin{equation}
\sqrt{\langle X^{2} \rangle -\langle X \rangle^{2}} \sim k_{B}T
\label{eq:def_temp}
\end{equation}
($k_{B}$: Boltzmann constant) 
to decrease to zero slowly enough 
during the Markov Chain Monte Carlo method for 
a given problem (the cost function or Hamiltonian). 
Then, the energy of the system 
does not decrease monotonically and sometimes 
it increases to escape the local minima in the energy landscape. 
We should keep in mind that 
the fluctuation vanishes at zero temperature as shown in the definition (\ref{eq:def_temp}).  
However, even at zero temperature, 
another type of fluctuation could be observed. 
Namely, 
we have the well-known uncertainty principle 
for non-commutative operators 
$\hat{X},\hat{Y}$ as 
\begin{equation}
\sqrt{\langle (\Delta \hat{X})^{2} \rangle 
\langle (\Delta \hat{Y})^{2} \rangle} \geq 
\frac{1}{2}
\left|
\langle [\hat{X},\hat{Y}] 
\rangle
\right|
\end{equation}
where we defined $\Delta \hat{X} \equiv \hat{X}-\langle \hat{X} \rangle$, 
and $\langle \hat{X} \rangle$ denote an expectation for 
state vector $|\psi \rangle$ as $\langle \hat{X} \rangle \equiv  \langle \psi |\hat{X}|\psi \rangle$. 
This means that we cannot determine the physical quantities 
whose operators are non-commutative at the same time. 
Therefore, the uncertainty is nothing but fluctuation which survives even at the grand state ($T=0$).

Obviously, 
for some sorts of materials such as spin glasses, 
it is quite important for us to figure out the properties at low temperature 
because the grand state structures of energy surface are non-trivial due to frustration. 
Hence, the SA is very effective to clarify the 
properties of materials. At the same time, 
the SA has been widely used in various information 
engineering problems. 
Actually, most of problems conceding in those research fields 
can be described by random spin systems which cause 
frustration at low temperature.

Then, we have the following relevant issues to be clarified. 
\begin{itemize}
\item
Existence of replica symmetry breaking at low temperature 
under quantum fluctuation. Namely, 
at low temperature, whether replica symmetry breaking 
which is a typical feature of spin glasses might occur or not  when we add the quantum 
fluctuation to the system? 
\item
Is it possible to construct effective algorithm using 
equilibrium state instead of ground state to be obtained by quantum adiabatic evolution? 
\item
Infinite range models are solved if we accept to use the so-called 
static approximation in the thermodynamic limit. 
However, the gap decreases to zero at the critical point, 
which means that adiabatic evolution across the critical point 
is impossible to be described analytically even in the infinite range model. 
Then, is there any advantage to discuss the dynamics of 
infinite range models across the critical point? 
\end{itemize}
All these queries are concerning the infinite range Ising model 
put in a transverse field. 
In this paper, 
we revisit the model and reconsider the problems. 

This paper is organized as follows. 
In Sec. \ref{sec:model}, 
we introduce the infinite-range spin glass model and 
show how the model is used for the problem of computer science. 
In Sec. \ref{sec:Trans}, 
we introduce quantum fluctuation as a transverse field. 
In this section, we also mention about QA in the context of optimization. 
In Sec. \ref{sec:Equi}, 
we show the equivalence between 
thermal and quantum fluctuations in information processing. We will see that 
the validity of static approximation is crucial to justify many analytic results.  
In Sec. \ref{sec:dyn}, 
we show the dynamical equations with 
respect to a few order parameters could be obtained for the infinite-range model. 
The last section is summary. 
\section{Infinite-range disordered Ising spin models and computer science}
\label{sec:model}
In this section, we explain several problems in computer science 
could be formulated by means of infinite-range disordered Ising models, 
which are in general described by 
\begin{equation}
H = -\sum_{i,j}J_{ij}\sigma_{i}\sigma_{j} 
-h \sum_{i}\tau_{i} \sigma_{i} 
\label{eq:general_Hamiltonian}
\end{equation}
where $\sigma_{i}$ stands for an Ising spin 
taking $\sigma_{i}=\pm 1$, 
and the summation with respect to $i,j$ should be taken for 
all possible pairs (infinite-range). 
Usually, we choose 
the strength of interaction $J_{ij}$ as 
of order $1/N$ to make the total energy of the system extensive. 
In following, we will briefly show some such examples.
\subsection{Associative memories in neural networks}
The so-called Hopfield model \cite{Hopfield} 
which can explain associative memories in artificial brain is described by the Hamiltonian 
of type  (\ref{eq:general_Hamiltonian}) with 
$J_{ij}=\frac{1}{N} \sum_{\mu=1}^{P}\xi_{i}^{\mu}\xi_{j}^{\mu}$, 
where $\bm{\xi}^{\mu}=(\xi_{i}^{\mu},\cdots,\xi_{i}^{\mu}), \,
\mu=1,\cdots,p$ are 
memory patterns embedded in the network. 
Obviously, 
for random patterns whose components take 
$\xi_{i}^{\mu}=\pm 1$ randomly, 
the $J_{ij}$ becomes a Gaussian variable 
with zero mean and standard deviation $\sqrt{\alpha/N}$. 
As the result, the Hamiltonian (\ref{eq:general_Hamiltonian}) 
is identical to the Sherrington-Kirkpatrick model \cite{Sherrington}. 
Of course, 
in real brain, 
it is more likely to exist structural limitation 
for synaptic connection, namely, 
$J_{ij}$ might be rewritten 
as $J_{ij}=\frac{1}{N} \sum_{\mu=1}^{P}
w_{ij} 
\xi_{i}^{\mu}\xi_{j}^{\mu}$ using 
adjacent matrix $w_{ij}\in \{1,0\}$. 
However, in artificial neural network, we usually treat 
the case $w_{ij}=1,\,\forall_{(i,j)}$ as the first approximation. 
Then, the quality of retrieval of a single specific pattern $\bm{\xi}^{1}$ is measured by 
the overlap between the pattern and neuronal state vector $\bm{\sigma}=
(\sigma_{1},\cdots,\sigma_{N})$ as 
$m=\bm{\xi}^{1} \cdot \bm{\sigma}$. 
Then, we have two types of noise to prevent the system to 
retrieve the pattern. 
Namely, thermal noise surrounding neurons and 
cross-talk noise from the other embeddded patterns 
$\frac{1}{N} \sum_{\mu \neq 1} \sum_{j}\xi_{i}^{\mu}\xi_{j}^{\mu} \sigma_{j}$. 
The former is controlled by temperature and induces second-order phase transition, 
whereas the latter is enhanced by increase of the number of patterns and it 
causes a first-order phase transition which 
determines the critical capacity (loading rate) $\alpha_{c}=p_{c}/N$ \cite{AGS}. 
\subsection{Traveling salesman problem}
We next see the so-called traveling salesman problem (TSP) 
\cite{Vannimenus,Mezard,Mezard2,Uesaka}. 
A typical TSP is defined as 
to find the shortest path for a salesman to 
go around extensive number of cities $N$. 
Let us define 
the label of each city by $m=A_{1},A_{2},{\cdots},A_{N}$, 
and we define 
index $n$ as 
the order of his/her visit as 
$n=1,2,{\cdots},N$. 
Then, a specific path is given by 
a correspondence such as 
$f(n)=A_{m},\, (n=1,2,{\cdots},N, m=A_{1},A_{2},{\cdots},A_{N})$. 
Thus, the total distance for the salesman is now given as 
\begin{equation}
D = 
\frac{1}{2}\sum_{i=1}^{N}
\sum_{\alpha,\beta=A_{1}}^{A_{N}}
d(\alpha,\beta)t_{i,\alpha}(t_{i+1,\beta}+t_{i-1,\beta}) 
\end{equation}
where $d(\alpha, \beta)$ stands for the distance between 
cities $\alpha$ and $\beta$, and 
we also defined $t_{i,\alpha}=\delta_{\alpha, f(i)}$.  
Obviously, we need several constraints: 
\begin{itemize}
\item
He/She can not visit two distinct cities at 
the same time (each round $i$) 
\[
\sum_{i=1}^{N}
\sum_{\alpha=A_{1}}^{A_{N}}
\sum_{\beta \neq \alpha}t_{i,\alpha}t_{i,\beta}=0.
\] 
\item
He/She can not visit more than once for each city 
\[
\sum_{\alpha=A_{1}}^{A_{N}}\sum_{i=1}^{N}\sum_{j\neq i}
t_{i,\alpha} t_{j,\alpha}=0.
\] 
\item
He/She must visit each city only once 
\[
\sum_{i=1}^{N}\sum_{\alpha=A_{1}}^{A_{N}}
t_{i,\alpha}=N.
\] 
\end{itemize}
By introducing these constraints 
via Lagrange multipliers $K,L,M$  
and replacing binary variables $t_{i,\alpha}=\{1,0\}$ to 
the Ising spins by 
$t_{i,\alpha}=\frac{1}{2}(\sigma_{i,\alpha}+1)$, we have the cost (Hamiltonian) by 
\begin{eqnarray}
H & = & 
-
\sum_{i,j=1}^{N}
\sum_{\alpha,\beta =A_{1}}^{A_{N}}
J_{i\alpha,j\beta}\, 
\sigma_{i,\alpha}\sigma_{j,\alpha}
-
\sum_{\alpha=A_{1}}^{A_{N}}
h_{\alpha}
\sum_{i=1}^{N}
\sigma_{i,\alpha}+C
\label{eq4b:TSP_Ising}
\end{eqnarray}
with 
$J_{i\alpha,j\beta} \equiv 
-(1/8) [d(\alpha,\beta)({\delta}_{j,i+1}+{\delta}_{j,i-1}) 
+2K{\delta}_{i,j}(1-{\delta}_{\alpha,\beta}) +  2L{\delta}_{\alpha,\beta}(1-{\delta}_{i,j})+2M]$ 
and 
$h_{\alpha} \equiv  MN-(1/2) [
\sum_{\beta=A_{1}}^{A_{N}}
d(\alpha,\beta)+(K+L)(N-1)+MN^{2}]$,  
$C \equiv  MN^{2}-MN^{3}+(1/4) [
N
\sum_{\alpha,\beta=A_{1}}^{A_{N}}
d(\alpha,\beta)+(K+L)N^{2}(N-1)+MN^{4}]$. 

From those expressions, 
we find 
that interactions between spins are now 
randomly distributed to satisfy 
$J_{i\alpha,j\beta} <0$ and 
it means that the system is nothing but 
a variant of 
the random anti-ferromagnetic Ising model. 
Moreover, all spins are now connected each other 
via the interactions, hence, we easily notice that the model 
is now categorized into the so-called infinite range model. 
I should be noted that the infinite-range anti-ferromagnets put in a transverse field Ising model 
was recently investigated extensively by \cite{Bikas,Chandra}. 
\subsection{Bayesian inference}
%
Inference problems such as 
image restoration, error-correcting codes and 
CDMA multiuser demodulator are related to the spin systems 
described by (\ref{eq:general_Hamiltonian}) in terms of well-known Bayes formula: 
\begin{equation}
P(\bm{\sigma}|\bm{\tau},\bm{J}) = 
\frac{P(\bm{\tau},\bm{J}|\bm{\sigma})P(\bm{\sigma})}
{\sum_{\bm{\sigma}}
P(\bm{\tau},\bm{J}|\bm{\sigma})P(\bm{\sigma})}
\end{equation}
where the likelihood $P(\bm{\tau},\bm{J}|\bm{\sigma})$ describes 
the degraded process of information or 
transmission of original messages through noisy channel, 
and $\bm{\tau},\bm{J}$ are both 
outputs of the noisy channel as we will see later on 
(see \cite{Nishimori_txt} for the basics).  
\subsubsection{Image restoration}
Actually, for instance, 
each original pixel $\sigma_{i}$ flips to the opposite 
sign $\tau_{i}=-\sigma_{i}$ with probability ${\rm e}^{-h}/2\cosh (h)$ 
and remains the same sign $\tau_{i}=\sigma_{i}$ as 
${\rm e}^{h}/2\cosh (h)$, 
we have $P(\bm{\tau}|\bm{\sigma})
={\exp}(h\sum_{i}\tau_{i}\sigma_{i})/\{2\cosh(h)\}^{N}$. 
To retrieve the original pixel (restoring image), 
we assume that each pair of pixels inclines to take the same value, 
namely, the prior $P(\bm{\sigma})$ could be chosen as 
$P(\bm{\sigma}) = {\exp}(J \sum_{ij}\sigma_{i}\sigma_{j})/\sum_{\bm{\sigma}}
{\exp}(J \sum_{ij}\sigma_{i}\sigma_{j})$.  
Then, we have the Hamiltonian of image restoration 
\cite{NW,TanakaK} 
which is now defined by $H=-\log P(\bm{\tau}|\bm{\sigma})P(\bm{\sigma})$ as 
\begin{equation}
H = -J \sum_{ij}\sigma_{i}\sigma_{j} -h \sum_{i}\tau_{i}\sigma_{i}. 
\label{eq:image}
\end{equation}
This is nothing but the Hamiltonian for the random field Ising model. 
Usually, 
digital image is defined on two-dimensional square lattice, 
hence, (\ref{eq:image}) is just a substitute for analyzing the performance. 
Actually, analysis of the solvable model (\ref{eq:image}) 
gives us a good guideline for the performance and 
quantum and classical Monte Carlo simulations  
have suggested that the performance evaluated by the infinite-range model 
(\ref{eq:image}) is qualitatively the same as the two-dimensional image restoration \cite{Inoue_PRE}. 
We also should mention 
here that the above formulation is valid for 
the binary image, however, 
the extension for grayscale or color images is possible. 
Actually, very recently, Cohen {\it et. al.} 
proposed a formula to present colored images and videos 
using Ising-like model \cite{CohenE2013,CohenE2014}. 

As in the same framework, 
one can describe the problem of 
error-correcting codes or CDMA multiuser demodulator by the following effective Hamiltonians. 
\subsubsection{Error-correcting codes}
For Sourlas error-correcting codes \cite{Sourlas} (see also \cite{Kabashima} 
for LDPC codes which is more useful for practical purpose), we have 
\begin{equation}
H = -\sum_{i1<\cdots<ip}J_{i1,\cdots,ip}\, 
\sigma_{i1}\cdots \sigma_{ip}
\label{eq:ecc}
\end{equation}
where we transmit 
the product (parity) of $p\,(\leq N)$ original bits 
among the original message given by $N$-dimensional vector 
$\bm{\xi}=(\xi_{1},\cdots,\xi_{N})$. 
This means that we transmit ${}_{n}C_{p}$ number of parities through the channel.  
Then, 
each product (parity) is degraded by an additive white Gaussian noise $\delta \sim \mathcal{N}(0,1)$ as 
\begin{equation}
J_{i1,\cdots,ip} \equiv (J_{0} p!/N^{p-1}) (\xi_{11} \cdots \xi_{ip}) + \sqrt{J^{2}p!/2N^{p-1}} \delta, 
\end{equation}
where $J_{0}/J^{2}$ is signal-to-noise ratio. 
Hence, when there does not exist any noise ($J=0$), 
the estimate which minimizes the Hamiltonian (\ref{eq:ecc}) is 
apparently $(\sigma_{1},\cdots,\sigma_{N}) = \bm{\sigma}=\bm{\xi}$, however, 
for $J\neq 0$, the best possible estimate might not be one which minimizes (\ref{eq:ecc}) 
but some another one. 
\subsubsection{CDMA multiuser demodulator}
We have the following Hamiltonian: 
\begin{equation}
H = \frac{1}{2N}\sum_{i,j}
\sum_{k=1}^{K}
\eta_{i}^{k}\eta_{j}^{k}
\sigma_{i} \sigma_{j} 
-
\frac{1}{\sqrt{N}}
\sum_{i}\sum_{k=1}^{K}
\eta_{i}^{k}y^{k}\sigma_{i}
\label{eq:cdma}
\end{equation}
for CDMA multiuser demodulator \cite{Tanaka}. 
In CDMA, 
the base station receives 
signals from $N$ users as a product of 
original information $\xi_{i}$ and spread codes $\eta_{i}^{k}, \,k=1,\cdots,K$ 
as $y^{k}=(1/\sqrt{N}) \sum_{i=1}^{N} \eta_{i}^{k} \xi_{i} + \epsilon^{k}$ 
where $\epsilon^{k}$ is an additive white Gaussian noise 
with mean zero and variance $\beta^{-1}$. 
Then, our problem is to estimate 
$\bm{\xi}$ from the observable $\bm{y}$ for a 
given spread codes matrix $\bm{\eta}^{k}, k=1,\cdots,K$. 
When we use 
$\sigma_{i}$ as 
 the estimate of original information $\xi_{i}$ and 
 assume a uniform prior for $\bm{\sigma}$ as 
 $P(\bm{\sigma})=1/2^{N}$, we have the posterior 
 $P(\bm{\sigma}|\bm{y})=P(\bm{y}|\bm{\sigma})
  \propto {\exp}[-\beta \sum_{k=1}^{K}(y^{k}-(1/\sqrt{N})\sum_{i=1}^{N}
 \eta_{i}^{k}\xi_{i})^{2}]  = {\exp}(-\beta H)$. 
Then, we have the Hamiltonian (\ref{eq:cdma}). 

As we already mentioned, 
digital image should be defined on two-dimensional square lattice properly, 
hence, (\ref{eq:image}) is just a substitute for analyzing the performance, 
however, (\ref{eq:ecc}) and (\ref{eq:cdma}) are both regarded as infinite range models 
and the models can describe the system properly. 
\section{Transverse field as quantum fluctuation}
\label{sec:Trans}
In classical system, 
the noise in neuronal systems,  
optimization problems or 
inference problems 
is introduced by transition probability between 
states. We have a lot of ways to 
simulate the noise by using appropriate stochastic processes. 
For instance, we might choose 
the probability $P(\sigma_{i})$ as 
\begin{equation}
P(\sigma_{i}) = 
\frac{1}{2}[
1+\sigma_{i} \tanh (\beta h_{i})],\,\,h_{i}=\sum_{j}J_{ij}\sigma_{j}
\end{equation}
where $\beta$ stands for inverse temperature $\beta =(k_{\rm B}T)^{-1}$. 
We have $P(\sigma_{i})=\frac{1}{2}[1+\sigma_{i}{\rm sgn}(h_{i})]$ 
at zero temperature ($\beta \to \infty$),  
which means that the spin $\sigma_{i}$ takes $+1$ with unit probability 
$P(\sigma_{i})=\frac{1}{2}[1+{\rm sgn}(h_{i})]=1$ 
when the local field $h_{i}$ surrounds a spin is positive $h_{i}>0$ 
and vice versa. 

On the other hand, when we recast the ingredients of the system 
from spin $\sigma_{i}$ to the Pauli matrix $\hat{\sigma}_{i}^{z}$ and 
unit matrix $\hat{I}$ as 
\begin{equation}
\hat{\sigma}_{i}^{z} = 
\left(
\begin{array}{cc}
1 & 0 \\
0 & -1
\end{array}
\right), \,\,\,
\hat{I}  = 
\left(
\begin{array}{cc}
1 & 0 \\
0 & 1
\end{array}
\right), 
\end{equation}
we have another expression of Hamiltonian as a large matrix with size $2^{N} \times 2^{N}$ as 
$\hat{H}_{0} \equiv  
-\sum_{ij}J_{ij} 
(\hat{I}^{\otimes (i-1)} \hat{\sigma}_{i}^{z} \hat{I}^{\otimes (N-i)}
)
(\hat{I}^{\otimes (j-1)} \hat{\sigma}_{i}^{z} \hat{I}^{\otimes (N-j)}
)
-h\sum_{i}\tau_{i} 
(\hat{I}^{\otimes (i-1)} \hat{\sigma}_{i}^{z} \hat{I}^{\otimes (N-i)})$. 
Then, in terms of optimization, the problem is rewritten so as to find the lowest 
energy state $|\psi \rangle = |\pm \rangle_{1} \otimes \cdots \otimes |\pm \rangle_{N}$ among all possible $2^{N}$ states.  
This is purely classical problem and we can use SA to achieve it. 
However, when we put the term 
$\hat{H}_{1}=\sum_{i}
\hat{I}^{\otimes (i-1)} \hat{\sigma}_{i}^{x} \hat{I}^{\otimes (N-i)}$
to the original (classical) Hamiltonian $\hat{H}_{0}$. 
Then, due to that fact that matrices $\hat{\sigma}_{i}^{x}$ and $\hat{\sigma}_{i}^{z}$ are non-commutative, 
the eigenstate $|\pm \rangle_{i}=\mbox{}^{t}(1,0),\mbox{}^{t}(0,1)$ satisfying $\hat{\sigma}_{i}^{z}| \pm \rangle_{i} = \sigma_{i}| \pm \rangle_{i}, 
\sigma_{i}=\pm 1$, 
the following single qubit flip is induced. 
\begin{equation}
\hat{\sigma}_{i}^{x} |+ \rangle_{i}= 
\left(
\begin{array}{cc}
0 & 1 \\
1 & 0 
\end{array}
\right)
\left(
\begin{array}{c}
1\\
0
\end{array}
\right)=
\left(
\begin{array}{c}
0\\
1
\end{array}
\right)=|-\rangle_{i},\,\,
\left(
\begin{array}{cc}
0 & 1 \\
1 & 0 
\end{array}
\right)
\left(
\begin{array}{c}
0\\
1
\end{array}
\right)=
\left(
\begin{array}{c}
1\\
0
\end{array}
\right)
\end{equation}
This sort of single qubit flipping causes 
quantum fluctuation even at the grand state. 
Hence, when we construct full Hamiltonian:  
\begin{equation}
\hat{H}=\hat{H}_{0}+\Gamma \hat{H}_{1},\,\,
\mbox{or}\,\,\,
\hat{H} = 
\frac{t}{\tau} \hat{H}_{0}+ 
\left(
1-
\frac{t}{\tau}
\right) \hat{H}_{1}
\end{equation}
and schedule the parameter as 
$\Gamma \to 0$ or $t \to \tau$ keeping the eigenstate (grand state) adiabatically for each time step, 
we obtain the grand state of the original (classical) Hamiltonian $\hat{H}_{0}$. 
This procedure is referred to as QA.  
%
\section{Equivalence between thermal and quantum fluctuations}
\label{sec:Equi}
%
In the Hopfield model, the equivalence of two kinds of fluctuation, 
namely, thermal and quantum which are controlled by 
$T$ and $\Gamma$,  was discussed by Nishimori and Nonomura \cite{Nishimori1996}. 
They drew the phase diagrams 
$\alpha$-$T$ and $\alpha$-$\Gamma$ and found that 
these phase diagrams are qualitatively the same which means 
that two distinct fluctuations are almost equivalent 
from the viewpoint of thermodynamics.  
In the Hopfield model, 
the fluctuation is a sort of `noise' which 
prevent the network to retrieve a specific embedded pattern. 
Hence, the above evidence tells us 
that as the origin of noise in artificial brain, 
one can consider the both thermal and quantum mechanism. 
It might be important for us to check 
the equivalence for much more practical use of fluctuation. 
%
\subsection{The Suzuki-Trotter formula and static approximation}
For inference, as an estimate $\bar{\xi}_{i}$ of the original bit 
$\xi_{i}$ which is degraded by some noise as $\tau_{i}$, 
we might use the expectation of a single qubit over the 
density matrix $\hat{\rho}={\rm e}^{-\hat{H}}/{\rm tr} \, {\rm e}^{-\hat{H}}
$ in terms of the Hamiltonian $\hat{H}$. 
Namely, we might use 
\begin{equation}
\overline{\xi}_{i} = {\rm sgn}
\left[
{\rm tr} (\hat{\sigma}_{i}^{z} \hat{\rho})
\right]. 
\end{equation}
In the classical limit $\Gamma \to 0$ at finite temperature 
$T=1$, the above estimate is optimal in the context of Bayesian statistics 
on the so-called Nishimori line at which 
the macroscopic variable such as noise amplitude or model parameters 
appearing in the prior distribution are identical to the corresponding true values \cite{Nishimori_txt}. 
However, it is not trivial if a similar condition to the Nishimori line is also observed 
even at the grand state $T=0$ with a finite amplitude of transverse field $\Gamma$. 
As we saw already, 
when we consider the wave function: 
\begin{equation}
|\psi (m) \rangle = |\pm \rangle_{1} \otimes \cdots \otimes |\pm  \rangle_{N} 
\end{equation}
where 
$|\pm \rangle_{i}, i=1,\cdots,N$ diagonalizes the classical part of the Hamiltonian, that is, 
$\hat{\sigma}_{i}^{z}
|\pm \rangle_{i} = \sigma_{i}
|\pm \rangle_{i}$,  
the density matrix is rewritten as 
\begin{equation}
\hat{\rho} = 
\frac{
w_{m} |\psi (m) \rangle \langle \psi (m) | 
\prod_{i} {\rm e}^{\Gamma \hat{\sigma}_{i}^{x}} 
|\psi (m) \rangle 
\langle \psi (m)|}
{\sum_{m=1}^{2^{N}}
w_{m} |\psi (m) \rangle \langle \psi (m) | 
\prod_{i} {\rm e}^{\Gamma \hat{\sigma}_{i}^{x}} 
|\psi (m) \rangle 
\langle \psi (m)|},
\end{equation}
where $w_{m}$ denotes the classical Boltzmann factor 
${\exp}[\sum_{ij}J_{ij}\sigma_{i}\sigma_{j}]$, namely, 
the diagonal components of the $2^{N} \times 2^{N}$ matrix 
$\langle \psi (m)| {\exp}[{\sum_{ij}J_{ij}\hat{\sigma}_{i}^{z}
\hat{\sigma}_{j}^{z}}]| \psi (m) \rangle$. 
The off-diagonal part 
$ \langle \psi (m) | 
\prod_{i} {\exp}[\Gamma \hat{\sigma}_{i}^{x}] 
|\psi (m) \rangle$ induces tunneling between states $|\psi (m) \rangle$. 
In general it is very hard to diagonalize the above large size matrix, 
and we usually use the following Suzuki-Trotter decomposition \cite{Masuo_Suzuki}: 
\begin{eqnarray}
{\rm tr}\, {\exp}
\left(
\hat{A}+\hat{B}
\right) & = & 
\lim_{P \to \infty} 
{\rm tr}
\left(
{\exp}
(
\hat{A}/P
)\, 
{\exp}
(
\hat{B}/P
)
\right)^{P}
\label{eq:ST}
\end{eqnarray}
for non-commutative 
operators $\hat{A}$ and $\hat{B}$. 
Then, the estimate $\bar{\xi}_{i}$ is calculated in terms of trace in the classical system 
whose dimension increases by $1$. 
Especially, we should notice that 
for the infinite range model, analytical evaluation of the estimate is 
possible because it is basically described by a single qubit problem 
in the limit of $N \to \infty$.  

In fact,  after ST formula (\ref{eq:ST}) for pure ferromagnetic Ising model $J_{ij}=J,\,\forall (i,j)$ 
in a transverse field, we typically encounter the following calculation. 
\begin{equation}
\bar{\xi}_{i} = 
\lim_{P\to \infty} 
\frac{{\rm tr}_{\{\sigma\}}
\sigma (k) \exp [\frac{\beta J}{P}\sum_{l=1}^{P}m_{l}\sigma (l) + B\sum_{l=1}^{P}\sigma (l)\sigma (l+1)]}
{{\rm tr}_{\{\sigma\}}
\exp  [\frac{\beta J}{P}\sum_{l=1}^{P}m_{l}\sigma (l) + B\sum_{l=1}^{P}\sigma (l)\sigma (l+1)]}  
\label{eq:local_mag}
\end{equation}
where $l$ denotes the Trotter slice and 
we should notice that 
the above quantity could be regarded as 
a local magnetization for one-dimensional Ising chain 
having spin-spin interaction with a strength $B=(1/2) \log \coth (\beta \Gamma/P)$ and 
local filed on cite $m_{l}$. 
To proceed the calculation, we usually use the so-called static approximation \cite{Thirumalai}, 
which means the local field $m_{l}$ is independent of 
the site $l$, namely, $m_{l}=m$. 
Then, we utilize the inverse process of ST decomposition, one can 
evaluate the above quantities as a single qubit problem. 
Of course, we might treat the above expectation 
as an Ising chain with a local field, 
however, the $l$-dependence of the field $m_{l}$ is 
non-trivial and we do not have any idea, in particular, for disordered 
quantum spin systems. 

Although several meaningful approaches have been reported \cite{Obuchi},  
however, 
due to the absence of alternative of static approximation, 
the existence of replica symmetry breaking in the infinite range 
spin glasses put in a transverse field has been unsolved. 
Ray {\it et al.}  \cite{Ray} also attempted to draw the Almeida and Thouless line \cite{AT} by using Monte
Carlo simulations, and they found that it might be possible to
conclude that there is no replica symmetry breaking due to the
quantum tunneling effects even in the low-temperature regime, 
however, any remark 
that decides the argument still remains unsolved.

\subsection{Several case studies}

As we already mentioned, 
the equivalence between temperature and amplitude of transverse field 
is practically important in the literature 
of information processing, and we can show
the amplitude $\Gamma$-dependence of 
the performance measure by analytical arguments. 
As we discussed in the previous subsection, unfortunately, such analytic treatment is not exact 
for most of the cases, however, the result might provide us a good guide to 
analyze the performance of the algorithm. 
As we already pointed out, 
error-correcting codes of Sourlas type are 
described by spin glass with $p$-spin interactions and 
the model put in the transverse field was already investigated by \cite{Gold1,Gold2}. 
However, recently it was revisited to investigate the thermodynamic 
properties from the computer scientific point of view \cite{Inoue2005,Inoue2009,Otsubo}. 

\begin{figure}[ht]
\begin{center}
\includegraphics[width=12cm]{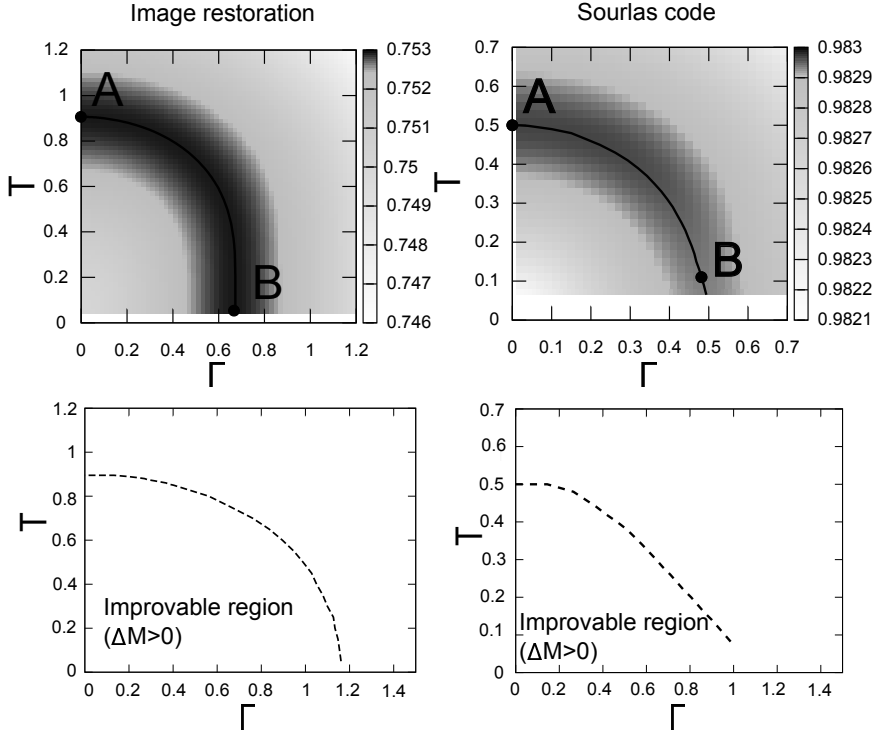}
\end{center}
\caption{\footnotesize  
Upper two panels show the value of overlap $M$ as a function of controlled parameters $T$ and $\Gamma$ 
 for image restoration (left) and Sourlas code (right). 
Overlap is maximized on solid lines. Lower two panels indicate improvable $(\Delta M >0$, see the definition of 
$\Delta M$ in (\ref{eq:def_DelM}) and worsened 
regions ($\Delta M<0$) for image restoration (left) and Sourlas code (right). $\Delta M=0$ holds on dashed lines. 
We set the parameters to $\tau_0=\tau=1.0, \;h=0.9,$, and $\beta_0=0.9$ for the image restoration model 
and $J=J_0=1.0,$ and $p=3$ for the Sourlas code (These were taken from ours \cite{Otsubo,Inoue_PRE}).
}
\label{fig:result_two}
\end{figure}
\begin{figure}[h]
\begin{center}
\includegraphics[width=12cm]{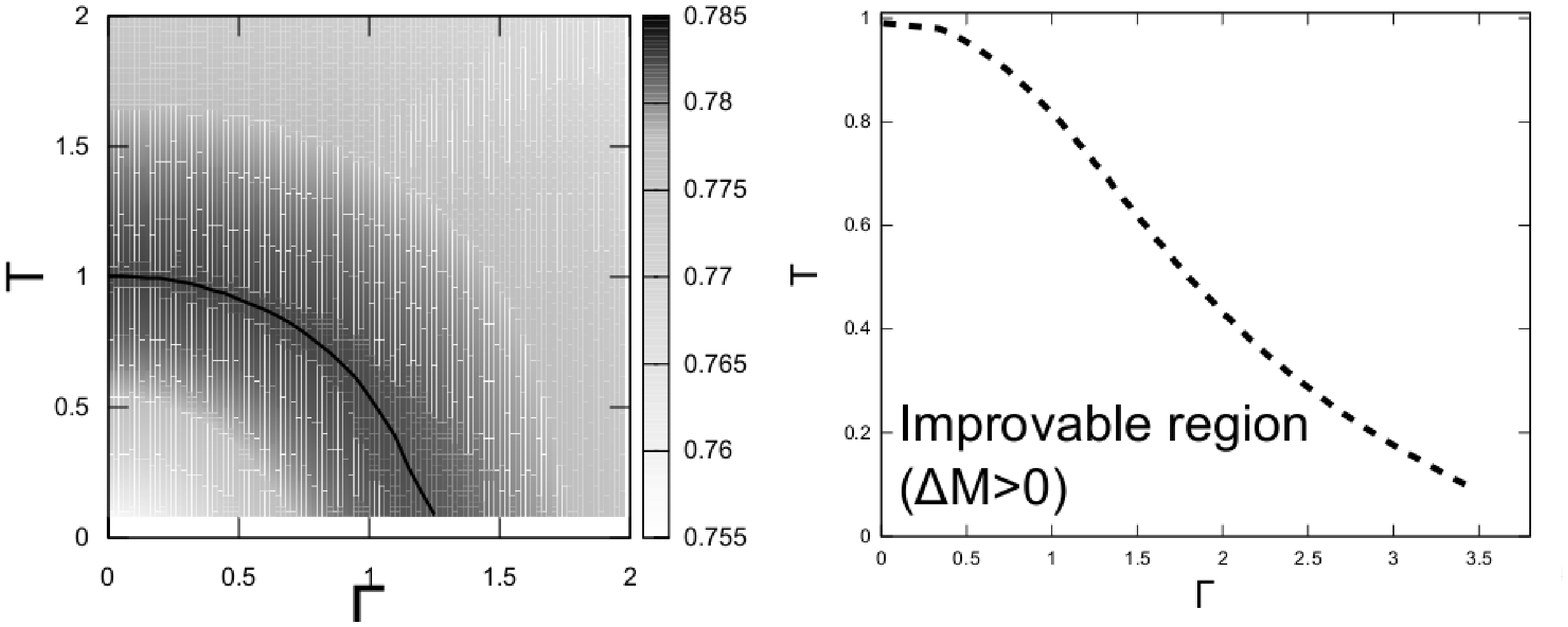}
\caption{\footnotesize 
Overlap $M$ on $\Gamma$-$T$ plane for $T_0=1.0$ and $\alpha=2.0$ (left) 
for CDMA multiuser demodulator.  
Solid line indicates location of peak in $M$. 
The right panel shows 
Improvable regions ($\Delta M>0$) in $T$-$\Gamma$ (left) $T_{0} = 1.0$. 
Dashed lines indicate border at which $\Delta M=0$ 
holds for $\alpha=K/N=2$ (These were taken from ours \cite{Otsubo2}). 
}
\label{fig:Mtop_Gamma}
\end{center}
\end{figure}

To quantify the performance,  
here we use the overlap: 
\begin{equation}
M = 
\frac{1}{N}
\sum_{i} \xi_{i} \overline{\xi}_{i}
\label{eq:def_DelM}
\end{equation}
and show the $\Gamma$-$T$ diagram with gradation in Fig.\ref{fig:result_two}
 (upper two panels). 
 We also display the critical line on which improvable (from estimation using conventional thermal (classical) fluctuation, 
 $\Delta M > 0$) 
 and worsened regions are separated 
 in Fig.\ref{fig:result_two} (lower two panels), 
 where we introduced the following quantity:
\begin{eqnarray}
\label{eq:delta_M}
\Delta M (\Gamma,\Omega)=M(\Gamma,\Omega)-M(0,\Omega),
\end{eqnarray}
that is, the difference between the overlaps at $\Omega=\{T_0,T,\alpha \equiv K/N\}$ 
with and without the quantum fluctuations. 
 The solid lines in the upper two panels indicate the lines on which overlap is maximized. 
 A peak appears in the overlap and then the location of the peak is roughly the same as that in the results
 for the classical case.  
 We also plot the same quantities as in Fig. \ref{fig:result_two} 
 for CDMA multiuser demodulator in Fig. \ref{fig:Mtop_Gamma}. 
\mbox{}

From these figures, 
we might conclude that 
the equivalence of two distinct fluctuations 
is observed even in the applications to various problems in computer science.  
\section{Dynamics, quantum phase transitions, across critical point}
\label{sec:dyn}
As we already mentioned, 
quantum Monte Carlo method is very powerful, 
however, when we simulate the quantum system at zero temperature in which quantum effect is essential, 
we encounter some technical difficulties. 
Therefore, it might be useful to look for exactly (analytically) solvable 
model to investigate the dynamical properties without computer simulations. 
Here we show that we can describe the 
quantum Monte Carlo dynamics by means of macroscopic order parameter 
flows for some class of infinite range models \cite{Inoue2010}. 
Note that such order parameter flows for classical Monte Carlo dynamics 
were already obtained in \cite{Coolen1988}. 
We also should note that 
Bapst and Semerjian recently extended the mean-field quantum dynamics \cite{Bapst}. 

Here we consider the pure ferromagnetic Ising model 
put in a transverse field. 
In terms of quantum Monte Carlo method, 
the master equation 
for the probability of the microscopic states  
on the $k$-th Trotter slice $p_{t} (\mbox{\boldmath $\sigma$}_{k})$  
is written by 
\begin{eqnarray}
\frac{dp_{t}(\mbox{\boldmath $\sigma$}_{k})}
{dt} & = & 
\sum_{i=1}^{N}
\left[
p_{t}(F_{i} ^{(k)}(\mbox{\boldmath $\sigma$}_{k}) )
w_{i}(F_{i}^{(k)}(\mbox{\boldmath $\sigma$}))-
p_{t}(\mbox{\boldmath $\sigma$}_{k})
w_{i}(\mbox{\boldmath $\sigma$}_{k})
\right] \\
F_{i}^{(k)} (\mbox{\boldmath $\sigma$}_{k}) & \equiv & 
(\sigma_{1}(k),\cdots, -\sigma_{i}(k),\cdots,\sigma_{N}(k)) 
\label{eq:master}
\end{eqnarray}
where $p_{t}(\mbox{\boldmath $\sigma$}_{k})$ denotes a 
probability that the system in the $k$-th Trotter slice 
is in a microscopic state $\mbox{\boldmath $\sigma$}_{k}$ at time $t$. 

The probability that the system is described by the 
magnetizationon each Trotter slice 
$m_{k}= N^{-1} \sum_{i} \sigma_{i}(k)$ at time $t$ is given 
in terms of the probability $p_{t}(\mbox{\boldmath $\sigma$}_{k})$ for a given 
realization of the microscopic state as $P_{t}(m_{k})  =  
\sum_{\mbox{\boldmath $\sigma$}_{k}} 
p_{t}(\mbox{\boldmath $\sigma$}_{k}) 
\delta (m_{k}-m_{k}(\mbox{\boldmath $\sigma$}_{k}))$. 
After simple algebra, we have 
\begin{eqnarray}
\frac{d P_{t}(m_{k})}
{dt} & = & 
\frac{\partial}{\partial m_{k}}\{m_{k}P_{t}(m_{k})\}
-\frac{\partial}{\partial m_{k}}\{
P_{t}(m_{k}) \langle \sigma (k) \rangle_{path}
\}
\label{eq:dPmkdt2}
\end{eqnarray}
where $\langle \sigma (k) \rangle_{path}$ 
is exactly the same form as the right hand hand side of equation (\ref{eq:local_mag}).  
In order to obtain the deterministic equation of order parameter, we should use the static 
approximation $m_{k} = m, \,\forall (k)$.  
Then, equation (\ref{eq:dPmkdt2}) leads to  
\begin{eqnarray}
\frac{d P_{t}(m)}
{dt} & = & 
\frac{\partial}{\partial m}
\{
m P_{t}(m)
\}
-
\frac{\partial}{\partial m}
\left\{
P_{t}(m)
\frac{Jm}{\sqrt{(Jm)^{2}+\Gamma^{2}}}
\tanh 
\sqrt{(Jm)^{2}+\Gamma^{2}}
\right\} 
\label{eq:use0}. 
\end{eqnarray}
Finally, substituting the form $P_{t}(m)=
\delta (m-m(t))$ 
into (\ref{eq:use0}) 
and 
making the integral by part with respect to 
$m$ after multiplying itself $m$,  we obtain 
the following deterministic equation.  
\begin{eqnarray}
\frac{dm}{dt} & = & 
-m  + 
\frac{Jm}{\sqrt{(Jm)^{2}+\Gamma^{2}}}
\tanh \beta 
\sqrt{(Jm)^{2}+\Gamma^{2}} 
\label{eq:dyn}
\end{eqnarray}
It is easy to see that the steady state $dm/dt=0$ is nothing but 
the equilibrium state described by the equation of state 
$m  = 
Jm\{\sqrt{(Jm)^{2}+\Gamma^{2}}
\}^{-1}
\tanh \beta 
\sqrt{(Jm)^{2}+\Gamma^{2}}$. 

\begin{figure}[ht]
\begin{center}
\includegraphics[width=6.4cm]{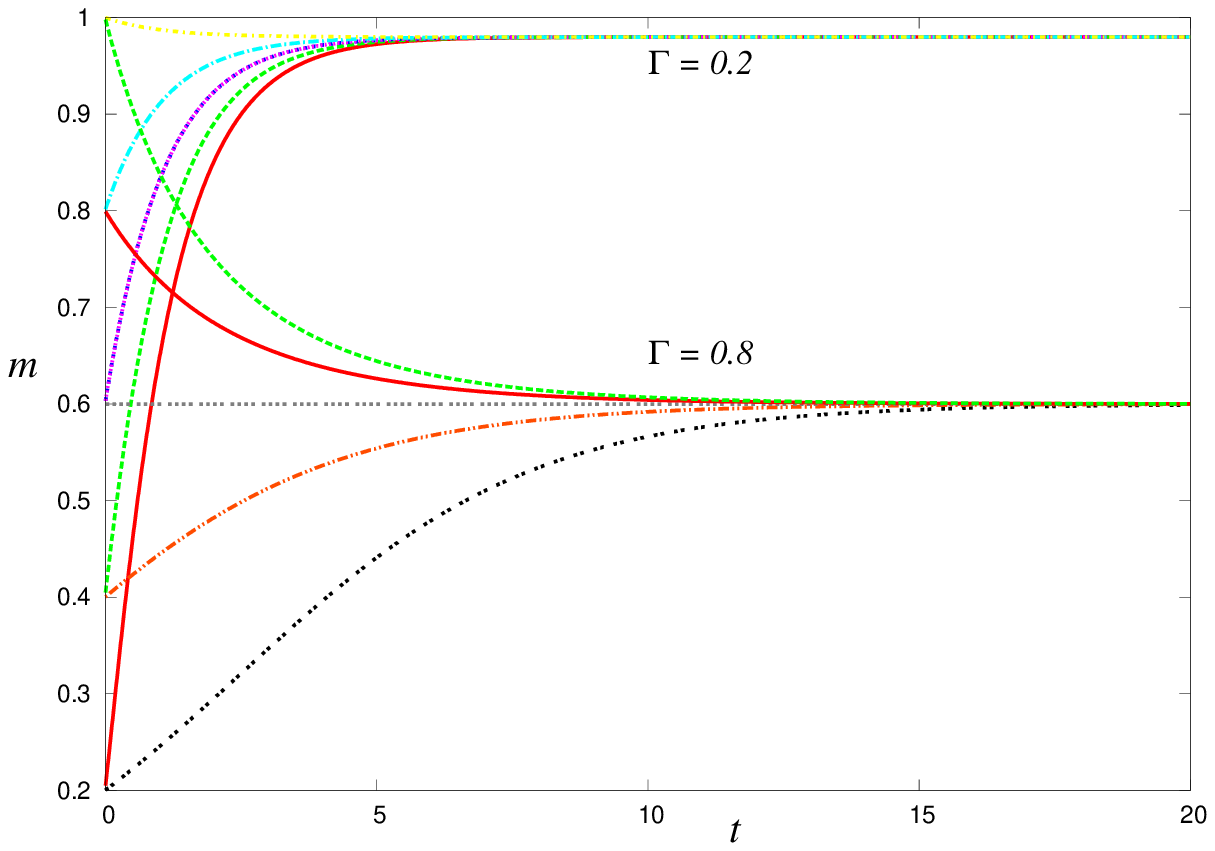}
\includegraphics[width=6.4cm]{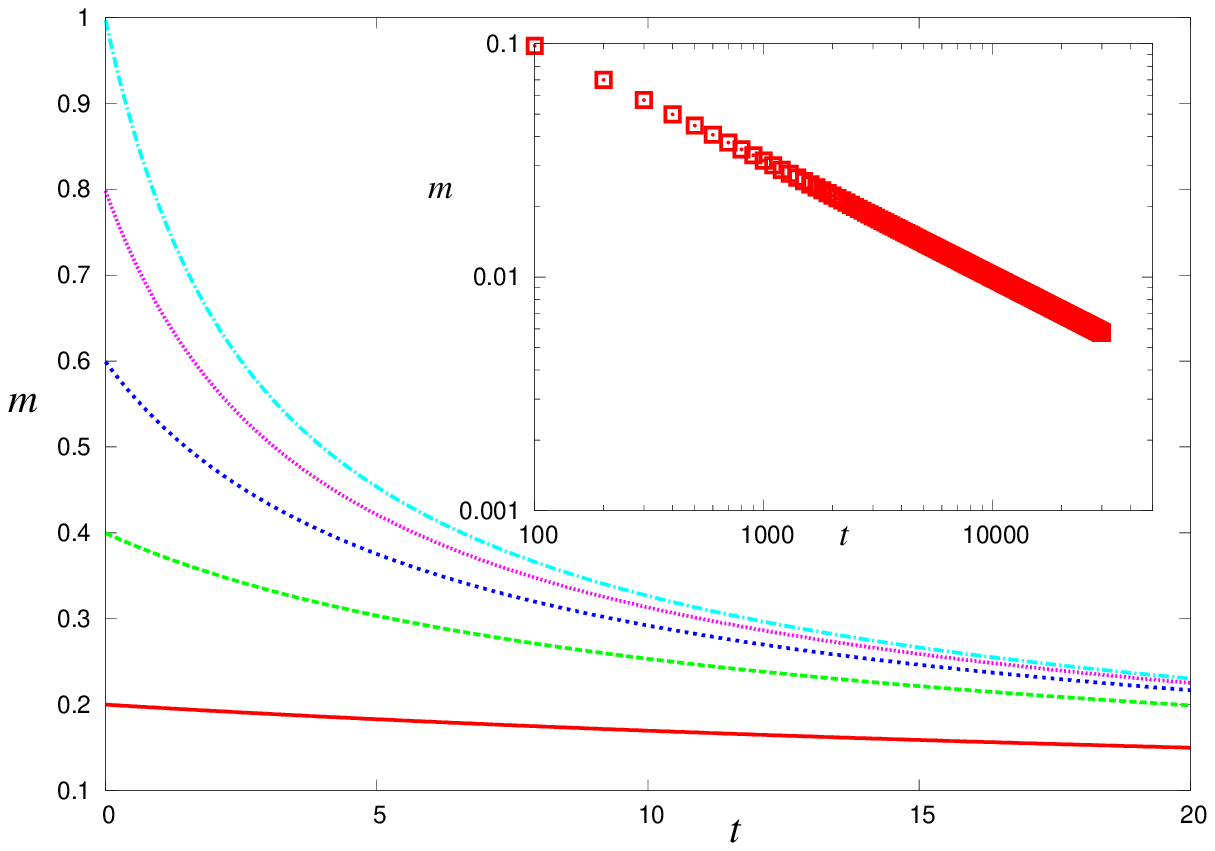}
\end{center}
\caption{\footnotesize 
Typical behaviour of zero-temperature 
dynamics described by (\ref{eq:dyn}) with $\beta =\infty$ 
far from the critical point $\Gamma_{c}=J=1$ of quantum phase transition (left). 
The right panel denotes 
the zero-temperature dynamics at the critical point. 
The inset shows the log-log plot of $m(t)$ indicating that 
the dynamical exponent in the critical slowing down is $\nu =1/2$ (These figures were taken from ours \cite{Inoue2010}). 
}
\label{fig:fg1}
\end{figure}
In Fig. \ref{fig:fg1} (left), 
we plot the typical behaviour of zero-temperature 
dynamics (equation (\ref{eq:dyn}) with $\beta =\infty$) 
far from the critical point $\Gamma_{c}=J=1$ of quantum phase transition. 
 We easily find that the dynamics exponentially converges to the steady state. 
The right panel denotes 
the zero-temperature dynamics at the critical point. 
The inset shows the log-log plot of $m(t)$ indicating critical slowing down $m(t) \simeq t^{-1/2}$.

\begin{figure}[ht]
\begin{center}
\includegraphics[width=6.4cm]{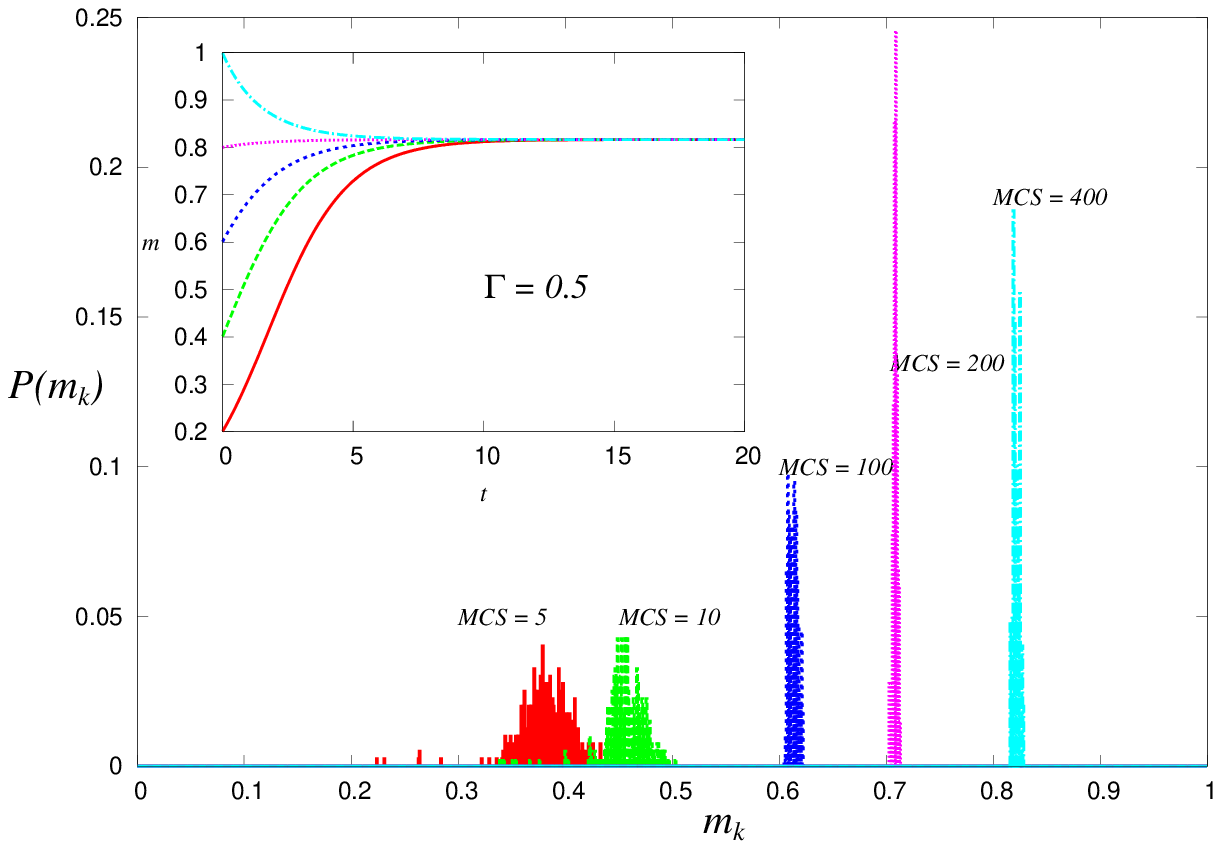}
\includegraphics[width=6.4cm]{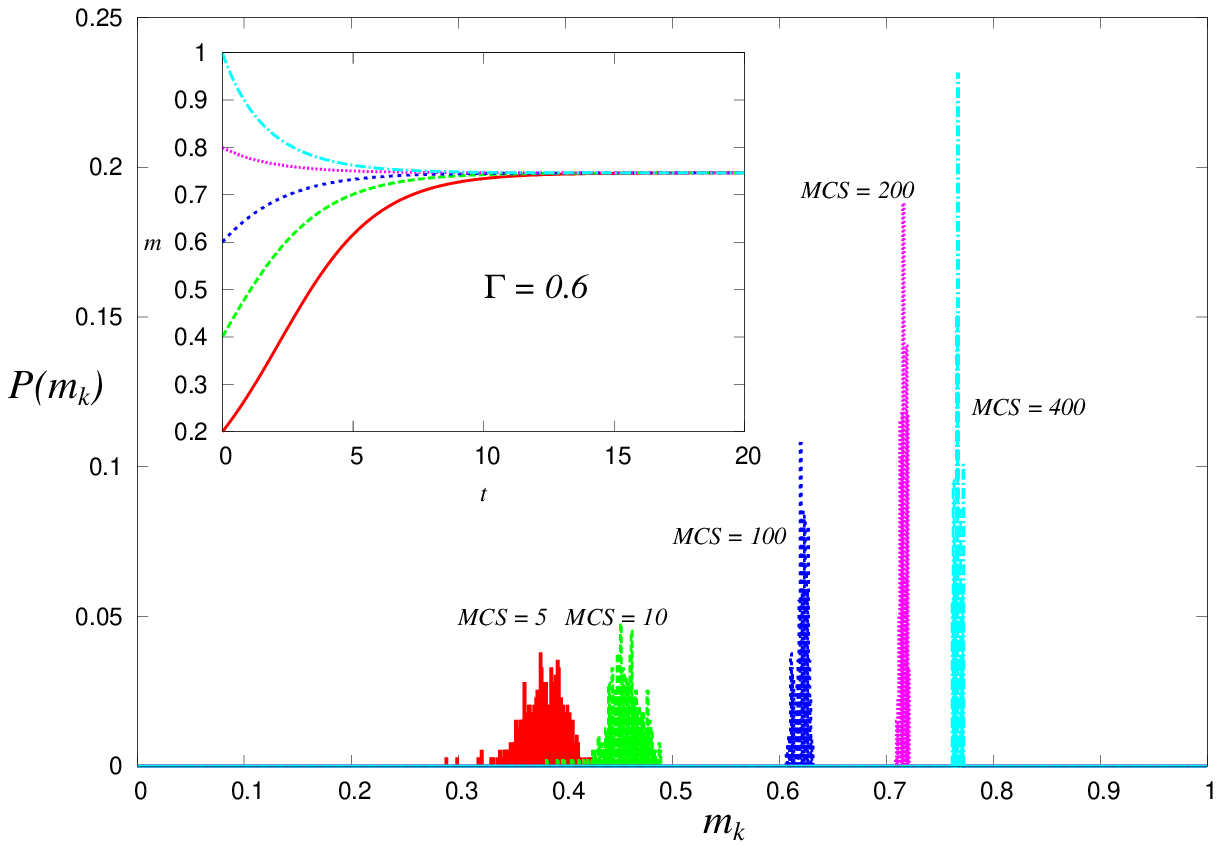}
\end{center}
\caption{\footnotesize 
Time evolution of the 
distribution $P(m_{k})$ calculated for 
finite size system with $N=M=400$. 
We choose the inverse temperature $\beta=2$ for 
$\Gamma=0.5$ (left) and $\Gamma=0.6$ (right). 
The inset in each panel 
denotes 
the deterministic flows of 
spontaneous magnetization 
calculated by (\ref{eq:dyn}) 
for corresponding parameter sets (These figures were taken from \cite{Inoue2010}). 
}
\label{fig:fg2}
\end{figure}

To confirm the validity of static approximation, we can carry out computer simulation for finite size system 
having $N=400$ spins. We observe the 
time evolving process of the 
histogram $P(m_{k})$ which is calculated from the $M=N=400$ copies of 
the Trotter slices. We show the result in Fig.  \ref{fig:fg2}. 
In this simulation, we chose the initial configuration in each 
Trotter slice randomly  and 
choose the inverse temperature $\beta=2$ for 
$\Gamma=0.5$ and $\Gamma=0.6$. 
From both panels in Fig.  \ref{fig:fg2}, 
we find that at the beginning, the $P(m_{k})$ is distributed due to the random set-up of the initial 
configuration, however, 
the fluctuation rapidly (eventually) shrinks to 
the delta function. 
After that,  the $P(m_{k})$ evolves as a delta function with the peak 
located at the value of spontaneous magnetization 
which is explicitly indicated in the inset of each panel. 
Hence, we are numerically confirmed that the static approximation 
is valid for the pure ferromagnetic infinite range Ising model 
(see also \cite{Inoue2010} for random field Ising model). 

As we mentioned in introduction, 
the gap decreases to zero at the critical point \cite{Young}, which means that adiabatic evolution across the critical point 
is impossible to be described analytically even in the infinite range model. 
However, we may discuss 
scaling behavior across the critical point (quenching) ({\it e.g.} \cite{Chandran,Heyl})
using the infinite range model. 
\section{Summary}
In this paper, we showed the relationship between the probabilistic 
information processing and theory of quantum spin glasses 
through the analysis of the infinite-range model.  
In classical spin glass, 
the infinite range model, that is, 
Sherrington-Kirkpatrick model \cite{Sherrington} is an exactly solved with in 
the Parisi scheme \cite{Mezard2}. 
This fact is very powerful because 
a lot of problems concerning 
computer science could be described by the variant of 
the SK model. 
However, as we discussed, 
for quantum extension of the solvable spin glass put in a transverse filed, 
we do not have yet exact solution due to the absence of alternative way 
of the static approximation. 
Therefore, the existence of replica symmetry breaking at low temperature has  
not yet been cleared \cite{Ray}. 
The static approximation is also required when we discuss the dynamics 
of quantum Monte Carlo analytically. 
Overcoming of this difficultly might be addressed as 
the most important issue in this research field. 
\subsection*{Acknowledgements}
The author gratefully acknowledges his friends and colleagues 
Y. Otsubo, 
K. Nagata, 
M. Okada, 
S. Suzuki, Y. Saika, A. Das, 
A. Chandra, S. Dasgupta, 
P. Sen, A. Dutta, S. Sharma and 
B. K. Chakrabarti 
for collaboration on this research topic.  
This study was financially supported by Grant-in-Aid for Scientific Research (C) of
Japan Society for the Promotion of Science (JSPS) No. 2533027803, 
Grant-in-Aid for Scientific Research (B) No. 26282089, 
and Grant-in-Aid for Scientific Research on Innovative Area No. 2512001313.

\end{document}